\documentclass[12pt]{iopart}
\usepackage{bm}
\usepackage{psfrag}
\usepackage{multirow}
\usepackage{color}
\usepackage{wasysym}
\usepackage{epsfig}

\renewcommand{\thefigure}{S\arabic{figure}}
\renewcommand{\thesection}{S\arabic{section}}
\renewcommand{\thetable}{S\arabic{table}}
\begin{document}

\title{Supplementary information for ``Selective sweeps in growing microbial colonies''}

\author{Kirill S Korolev, Melanie J I M\"uller, Nilay Karahan, Andrew W Murray, Oskar Hallatschek and David R Nelson}

\begin{abstract}
This is supplementary information for arXiv:1204.4896, which also appeared in Physical Biology 9, 026008 (2012).
\end{abstract}


In this supplementary information, we first describe in section~\ref{SExperiments} the methods used in the main text. We then show that the macroscopic spatio-genetic patterns of spatial competitions are independent of microscopic details, using both experimental and theoretical arguments: In section~\ref{SBuddingPattern} we show experimentally that sector shapes apply also to microbes with different cell division patterns. In section~\ref{S_Insensitivity_model_parameters}, we show that the patterns predicted by our reaction-diffusion model are not sensitive to the microscopic model parameters. Next, we investigate the duration of the initial stage of sector formation in our experiments in section~\ref{SSectorEstablishment}. Finally, we detail the statistical analysis of differences between different ways to measure relative fitness in section~\ref{SStatisticalAnalysis}.

\section{Methods}\label{SExperiments}

\textbf{Numerical integration.} We used an explicit forward-time centered-space~(FTCS) finite-difference method (6-point stencil) on a square grid to numerically solve the reaction-diffusion partial differential equations describing our model in space and time~\cite{press:numerical_recipes}.

\textbf{Strains.} 
For the competition experiments, our reference \textit{Saccharomyces cerevisiae} strains were the prototrophic W303 strains yJHK102, yJHK111, and yJHK112. These strains all have the genotype MAT\textbf{a} \textit{bud4$\Delta$::BUD4(S288C) can1-100}, and differ only at the His3 locus, with \textit{HIS3}, \textit{his3Δ::prACT1-ymCitrine-tADH1:His3MX6} and \textit{his3Δ::prACT1-ymCherry-tADH1:His3MX6} for yJHK102, yJHK111, and yJHK112, respectively. Thus, yJHK102 is not fluorescent, the strain yJHK111 constitutively expresses a yellow fluorescent protein, and yJHK112 constitutively expresses a red fluorescent protein, which is pseudo-colored as blue in the microscopy images shown.

The cycloheximide-resistant strain yMM8 is derived from yJHK111 by replacing \textit{CYH2::cyh2-Q37E}, which confers resistance to the drug cycloheximide \cite{StockleinBock81}. The budding mutant yMM22 is also derived from yJHK111 by replacing \textit{bud4$\Delta$::KANMX6}. The advantageous mutant $\alpha F^{\rm R}$-E04 was isolated in a screen for $\alpha$-factor resistant mutants of the strain DBY15084 (W303 MAT\textbf{a} \textit{ade2-1 CAN1 his3-11 leu2-3,112, trp1-1 URA3 bar1$\Delta$::ADE2 hml$\alpha\Delta$::LEU2}) performed in Ref.~\cite{LangMurray09}. This mutant is non-fluorescent.

In the text, we refer to the strains yJHK102, yJHK111, and yJHK112 as the non-fluorescent, yellow fluorescent, and red fluorescent wild-type strain, respectively. Further, we call the strains $\alpha F^{\rm R}$-E04, yMM8, and yMM22 the advantageous sterile mutant, the advantageous cycloheximide resistant mutant, and the budding mutant. 

\textbf{Growth conditions.} We used 1\% agarose plates with CSM (complete synthetic medium as described in Ref.~\cite{BurkeStearns00}, except 2$\;$g of adenine and 4$\;$g of leucine were used). Strains were pre-grown in liquid CSM at $30^{\circ}$C in exponential phase for more than 12 hours. They were then counted with a Coulter counter and mixed in appropriate ratio of mutant:wild-type to obtain well-separated sectors. For competitions of the advantageous sterile mutant with the wild-ype, this ratio was 1:500. For competitions of the cycloheximide-resitant mutant with the wild-type, the ratio was 1:100, 1:200, and 1:500 for cycloheximide concentrations $<50$nM, in the range 50-90nM, and $>90$nM, respectively.   
This mix was then inoculated on agar plates that had dried for 2 days post-pouring. For circular colonies, drops of 0.5$\;\mu$l of the mix were pipetted on the plate. For linear colonies, we dipped a sterilized razor-blade into the cell mix and then gently touched the agar surface with the razor-blade. For colliding circular colonies, two drops of 0.5$\;\mu$l of the mix were pipetted on the plate with centers approximately $1\;\rm{mm}$ apart. The plates were then incubated at $30^{\circ}$ in a humidified box for 8 days and imaged with a Zeiss Lumar stereoscope. 

\textbf{Image analysis.} Data analysis was performed using MatLab R2010. In particular, colony and sector boundaries were detected using the edge function in MatLab R2010. 

\textbf{Sector analysis.} The analysis of sectors is illustrated in figures~15 and~16. Sector boundaries from edge detection were separated into two bounding ``arms'' by tracing backwards from the two outmost points towards the colony center. The sector arms were then plotted in Cartesian coordinates~$x(y)$ for linear inoculations and in log-transformed polar coordinates~$\phi[\ln(r)]$ for circular inoculations. Next, the average sector arm position was subtracted from each sector arm, in order to average out wobble in the growth direction of the sector. The resulting sector arms were then fitted with straight lines, in accordance with equations~(10) and~(12), for long times, i.e. for large~$y$ or~$r$. This procedure is equivalent to fitting the average of the upper and lower arms of the sector. 
We ensured that the fitting was done only to established sectors. Since different sectors splayed off into linear behavior at different times, the start point for the fit was determined using the following condition: The maximal number of points from the colony boundary towards the inoculum was fitted that still gave $r^2$ value of at least $0.995$. 
Here, $r^2$ is the square of the Pearson correlation coefficient, whose value very close to 1 indicates a very good linear correlation \cite{Weiss07}, implying that our theory describes the sector boundaries very well. For the linear expansions, this condition led to fits at a distance from the inoculant of roughly $y>3\;\rm{mm}$, corresponding to times $t>120\;\rm{h}$. For the circular expansions, the fit range was $r/R_0>4$, or $t>90\;\rm{h}$. 
We also fitted the sectors with the fixed cutoff of $y>3\;\rm{mm}$ (linear expansions) and $r/R_0>4$ (circular expansions), and obtained the same results for the relative fitnesses within error bars.

\textbf{Sector analysis for small nonzero fitness differences.}
For small nonzero fitnesses, $0<s<0.02$, some sector boundaries failed the stringent high-quality-of-fit criterion of $r^2>0.995$. We attribute this to the fact that, for small relative fitnesses, fluctuations caused by genetic drift and other noise sources have larger impact on the sector shape than for higher fitness differences. In addition, the establishment time for sectors become very large for small $s$, see the discussion in section~\ref{SSectorEstablishment}, so that some of the sectors might not yet have fully established on the experimental timescale. In consequence, we ignored sectors that failed the $r^2>0.995$ criterion.

\textbf{Sector analysis for zero fitness differences.}
In cases where sectors boundaries were indistinguishable from straight lines, leading to a relative fitness value $s$ equal to zero within error bars, all sectors failed to fulfill the criterion $r^2>0.995$. The reason is that, for zero fitness differences, sector boundaries are dominated by genetic drift and other sources of fluctuations. In this case, we therefore fitted these sectors for radii $r/R_0>4$ (radial expansions) and for distances $y>3\;\rm{mm}$ (linear expansions).

\textbf{Analysis of colliding colonies.} The analysis of colliding colonies is shown in figure~17. To obtain this figure, we used Matlab R2010 to detect the edges of the two colonies, and to then fit circles to each colony as well as to the boundary between the colonies. The selective advantage was calculated from the ratio $x_0/R_b$, see equations~(20) and~(21). The selective advantage can also be calculated from the distance $x_0$ or from the radius $R_b$ alone by solving equation~(20) or equation~(21), respectively, for the selective advantage $s$. All three methods gave the same results for the selective advantage within the measurement uncertainty. Although the observed standard deviations were small for this assay, we suspect that it might suffer from a systematic error. This error could result from our assumption of constant expansion velocity ratios, which might be somewhat inaccurate at the early stage of colony growth; see the inset in figure~14 in the main text.

\textbf{Expansion velocities.} The analysis of radial expansion velocities is shown in figure~14 for the circular expansions. Radii of each colony were plotted against time. Each radial growth curve was fitted with a straight line over the same times as for the sector analysis, i.e. for times larger than~$90\;\rm{h}$. The relative fitness was then calculated as the ratio of the average growth velocities (slopes). Analogously, for the linear expansions, the increase in extension in $y$-direction of each razor-blade inoculation was fitted for times larger than~$120\;\rm{h}$. The relative fitness was again calculated as the ratio of the average velocities.

\textbf{Error calculations.} 
In order to obtain the selective advantages~$s$ for plate assays shown in table~1, the respective experiment was done on 2-3 different batches of plates, with 4-10 replicates each times. We found that the expansion velocities and fitnesses determined from replicates on the same batch of plates were very similar, while expansion velocities from different batches of plates were significantly different, with $p$-values below~$0.03$ determined by an ANOVA F-test. This statistical hypothesis test compares the variance within and between sets of replicates in order to test whether the several sets have the same expected value of the measured quantity, and a small $p$-value indicates that the means of the sets are significantly different~\cite{Weiss07}. The difference between different batches could be due to different plate humidities. We therefore calculated the errors of the velocities and fitnesses using a random cluster bootstrapping algorithm suitable for clustered data~\cite{DavisonHinkley97}. In contrast, fitnesses calculated from sectors (both from linear and radial expansions) and from colony collisions on different batches of plates were not significantly different.

\textbf{Fitness in liquid culture.} The growth rates in liquid culture were determined by taking standard growth curves of cells pre-grown in exponential phase in CSM for more than 12 hours, using a Coulter counter. The relative fitness in direct competition in liquid culture was determined with a flow-cytometer-based competition assay as described in Ref.~\cite{LangMurray09}.

\textbf{Growth of \textit{Pseudomonas aeruginosa} colonies.}
See Ref.~\protect{\cite{korolev:neutral_expansions}} for the strain information and growth conditions.

\section{Neutral competitions of strains with different fluorescence}\label{SNeutral}

\begin{table}
\begin{tabular}{|l@{\;\;}|l@{\;\;}|l@{\;\;}|}\hline
Competing reference strains & yellow : non-fluorescent & yellow : red\\
Assay	 & Selective advantage $s$& Selective advantage $s$ \\\hline
Liquid culture competition	& -0.01 $\pm$ 0.01 (N=3)			& \phantom{-}0.00 $\pm$ 0.01 (N=3)\\
Radial expansion sectors 	& \phantom{-}0.00 $\pm$ 0.01 (N=97)	& \phantom{-}0.00 $\pm$ 0.01 (N=96)\\\hline
\end{tabular}
\caption{Expression of a fluorescent protein in our reference strains does not incur a fitness cost: There is no significant selective advantage in competitions of the yellow wild-type yJHK111 with the non-fluorescent wild-type yJHK102, or with the red wild-type yJHK112. Errors are standard deviations, and the number $N$ of replicates is given in parentheses.}
\label{tab:neutral}
\end{table}

Our fitness assay does not rely on any assumptions about the exact origin of the fitness difference between the strains. The assay, however, would be more valuable if we can show that the introduction of fluorescence does not significantly affect the fitness. We therefore measured the relative fitnesses of our three reference strains, that differ only in the expression of a fluorescent protein: the non-fluorescent strain yJHK102, the yellow strain yJHK111 (constitutively expresses mCitrine), and the red strain yJHK112 (constitutively expresses mCherry). 
We confirmed that these strains are neutral with respect to each other, as measured using the liquid culture fitness assay and the radial plate sector assay, see table \ref{tab:neutral}. This means that the expression of our fluorescent proteins does not incur a fitness cost. 

\begin{figure}
\includegraphics[width=10cm]{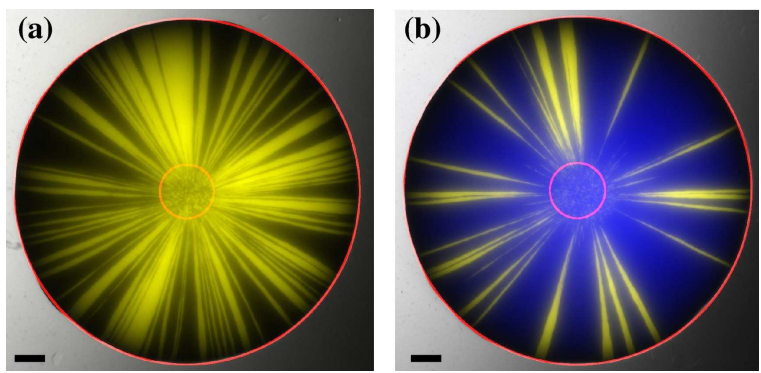} 
\caption{Neutral expansions of strains, that differ only in expression of a fluorescent protein, exhibit sectors with straight boundaries. 
(a) Competition of the yellow and non-fluorescent wild-type strains, inoculated in ratio 1:1.
(b) Competition of the yellow and red wild-type strains (pseudo-colored as blue for better contrast), inoculated in ratio 1:10.
Red circles indicate the size of the colony right after the inoculation and at the end of expansion. The scale bars are 1mm.}
\label{fig:neutral_radial_expansion}
\end{figure}

Examples of radial expansions of mixes of these strains are shown in figure \ref{fig:neutral_radial_expansion}. Note that the sector boundaries are straight lines, as predicted by equation (16) in the main text for $v_1=v_2$. An example of a linear expansion competition of the yellow and red reference strain is shown in the main text figure 1(a). The shape of neutral sectors was already discussed by Hallatschek et al., compare Fig. 4B in Ref.~\cite{HallatschekNelson:ExperimentalSegregation}.

\section{Insensitivity of sector shapes to details on cellular lengths scales}\label{SBuddingPattern}

Our theory describes the growth of a microbial colony, and the positions of boundaries of sectors and colony collisions, on spatial length scales much larger than the cell size. The predictions of our theory should therefore be independent of microscopic details on cellular length scales. In order to test this, we investigated whether the manner of cell division of \textit{S.~cerevisiae} influences the  macroscopic patterns observed in colony expansions. In addition, we conducted a competition experiment using the bacterium \textit{Pseudomonas aeruginosa}.

\subsection{Insensitivity of sector shapes to the \textit{S. cerevisiae} budding pattern}

Cells of \textit{S.\ cerevisiae} replicate by budding off daughter cells. The spatial locations for new buds are chosen non-randomly \cite{BartholomewMittwer53,ChantHerskowitz91}: Haploid cells bud in an axial pattern, wherein each bud emerges adjacent to the site of the previous bud, while diploid cells bud in a bipolar pattern, wherein successive buds can emerge from either end of the ellipsoidal mother cell. 

\begin{figure}
\includegraphics[width=\textwidth]{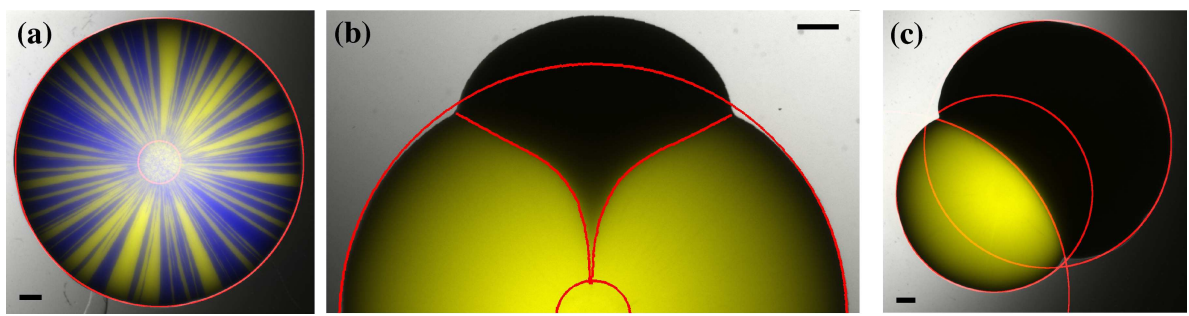} 
\caption{Different budding patterns do not change the sector shape: When replacing our yellow wild-type strain yJHK111 ($BUD4$ wild type, axial budding pattern) by the yellow budding mutant yMM22 ($bud4\Delta$, bipolar budding pattern), sector shapes remain the same.
(a) A radial expansion of the yellow budding mutant and the red wild-type yJHK112 (pseudo-colored as blue), inoculated at ratio 1:1, exhibits sectors with straight boundaries. This is the same assay as shown in figure~\ref{fig:neutral_radial_expansion}, but with the yellow wild-type replaced by the budding mutant. 
(b) Radial expansion and (c) colony collision of the yellow budding mutant and the non-fluorescent advantageous sterile mutant $\alpha$E04. These are the same assays as shown in the main text figures 16 and 17, but with the yellow wild-type replaced by the budding mutant.
All scale bars are 1mm.}
\label{fig:budding_mutants}
\end{figure}

\begin{table}
\begin{tabular}{|l@{\;\;}|l@{\;\;}|l@{\;\;}|}\hline
Assay 					 			& Selective advantage $s$\\ \hline
Liquid competition  	 			& $0.18 \pm 0.01 \;\, (N=3)$\\
Radial expansion sectors 			& $0.25 \pm 0.03 \;\, (N=11)$\\
Colony collisions					& $0.23 \pm 0.01 \;\, (N=12)$\\\hline 
\end{tabular}
\caption{Relative fitnesses of the advantageous sterile mutant $\alpha F^{\rm R}$E04 and the budding mutant yMM22.
Errors are standard deviations, and the number $N$ of replicates is given in parentheses. These relative fitnesses are very similar to the ones obtained for the competition of the sterile mutant with the yellow wild-type, see table 1 in the main text. }
\label{table:budding_mutants}
\end{table}

In order to test whether the budding pattern affects the formation of sector boundaries, we used \textit{haploid} strains with different budding patterns by modifying the \textit{BUD4} locus. Bud4p is a landmark protein required for the axial budding pattern in haploid cells: Haploids with \textit{BUD4} deletions exhibit the bipolar budding pattern \cite{ChantHerskowitz91}, but otherwise grow normally \cite{SandersHerskowitz96}. 
Our yellow wid-type strain yJHK111 has a functional BUD4 protein and exhibits the axial budding pattern. By deleting the $BUD4$ gene in this strain, we made the yellow budding mutant yMM22 with a bipolar budding pattern. We confirmed that the $bud4$ deletion is neutral by competing the budding mutant against the red wild-type yJHK112 in a liquid competition assay ($s=0.00\pm 0.01$, N=3). The budding mutant was also neutral in a radial expansion competition with the wild-type, forming sectors with straight boundaries ($s=0.00 \pm 0.01, N=65)$, see figure \ref{fig:budding_mutants}(a). The change in budding pattern therefore does not change the shape of neutral sectors.

We then repeated the direct competition experiments described in the main text for the advantageous sterile mutant $\alpha F^{\rm R}$E04 and the yellow wild-type yJHK111, but replaced the wild-type with the budding mutant. We only used competition assays with reasonable accuracy, i.e.\ the liquid culture competition, and the radial expansion sector and colony collision assays. Images of the latter two assays are shown in \ref{fig:budding_mutants}(b,c); they look very similar to the corresponding images in figures 16 and 17 of the main text. In particular, the sector and collision boundaries are described well by logarithmic spirals and circles, respectively, as predicted by our theory. 

The numerical values of the relative fitnesses of the advantageous sterile mutant and the budding mutant are listed in table~\ref{table:budding_mutants}. The relative fitnesses from the radial sector and the colony collision assays agree well with each other, but are larger than the relative fitness from liquid competition. The observed range of the relative fitnesses $s$ of the advantageous sterile mutant and the budding mutant is consistent with the relative fitnesses of the advantageous sterile mutant and the wild-type, compare table~1 in the main text.

\subsection{Sector shapes for the bacterium \textit{Pseudomonas aeruginosa}}

\begin{figure}
(a)\includegraphics[width=5cm]{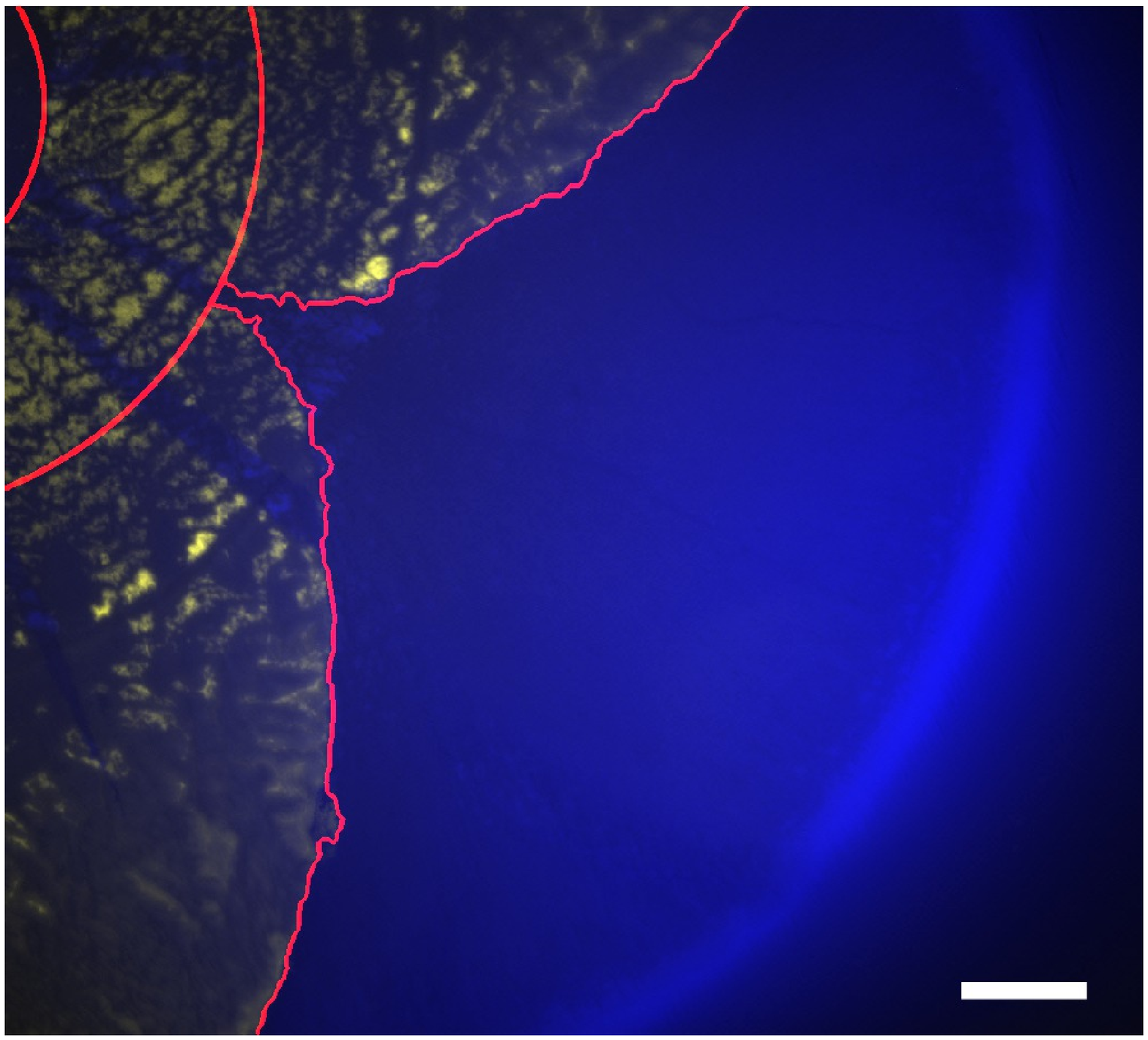}
(b)\includegraphics[width=6cm]{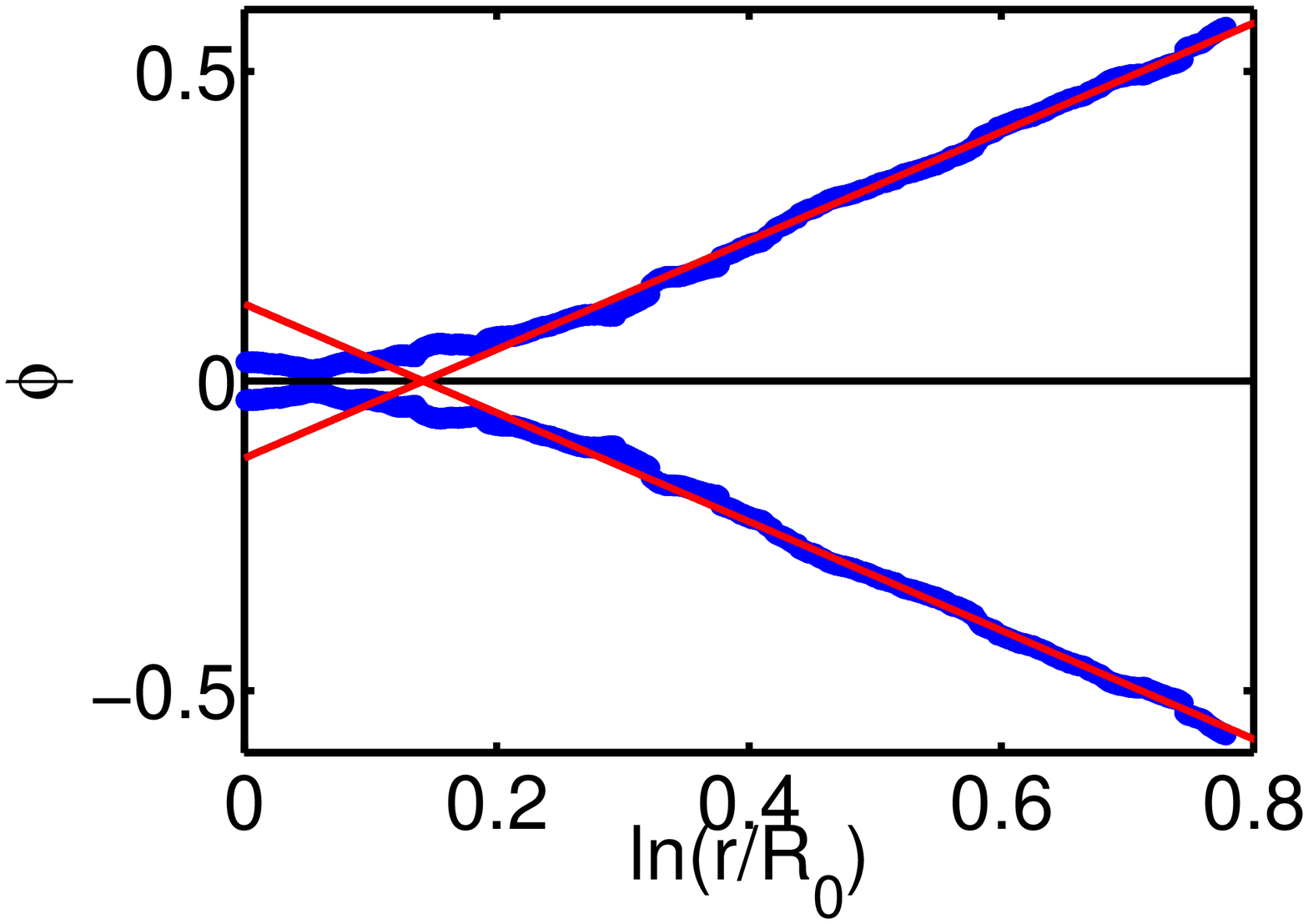}
\caption{Using the radial sectoring assay for the bacterium \textit{Pseudomonas aeruginosa}. 
(a) Fluorescent image of a spontaneous advantageous mutation (or another heritable phenotypic change) in a \textit{P.~aeruginosa} colony, which was grown from a circular drop of two neutral fluorescently labeled strains (yellow and blue). The smaller red circle marks the visible part of the homeland, while the larger red circle marks the approximate radius where the mutation occurred; its radius is taken as $R_0$. The initial sectoring is due to genetic drift, but the largest blue sector expands because the spontaneous mutation confers a selective advantage. The scalebar is 2mm.
(b) Sector boundaries (blue dots) extracted from (a), and the fits (red lines) to a logarithmic spiral. From the slope of the fit, we estimate that $s=0.33$. }
\label{FPaeruginosaCircular}
\end{figure}

To test our theory even further, we conducted a competition experiment in a different species, the bacterium \textit{Pseudomonas aeruginosa}. Figure~\ref{FPaeruginosaCircular}(a) shows the radial expansion of two neutral fluorescently labeled \textit{P.~aeruginosa} strains. A spontaneous advantageous mutation (or another heritable phenotypic change) gave rise to a large blue sector, whose boundaries can be well fitted by a logarithmic spiral, see figure~\ref{FPaeruginosaCircular}(b).

Given all the differences in growth and migration between the yeast \textit{S.\ cerevisiae} and the bacterium \textit{P.\ aeruginosa}, it is remarkable that the theoretically predicted logarithmic spiral describes the shapes of selective sweeps in both species so well.

In summary, all experimental evidence is this section supports our theoretical conclusion that spatio-genetic patterns created by range expansion and natural selection are rather insensitive to the microscopic details of the growth and migration.


\section{Robustness of reaction-diffusion model to microscopic details}\label{S_Insensitivity_model_parameters}

In this section, we present additional evidence that the macroscopic shape of spatial patterns depends only on~$v_{1}/v_{2}$ in our reaction-diffusion model represented by equation~(4) in the main text.

\begin{figure}
\includegraphics[width=6cm]{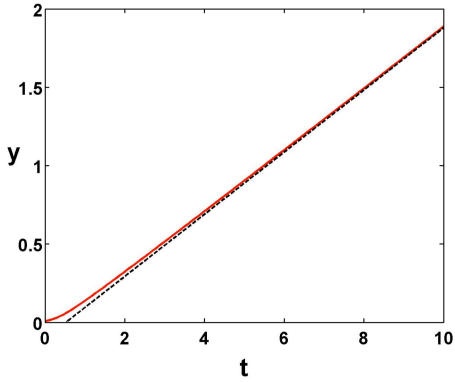} 
\caption{The velocity of colony expansion in the linear geometry quickly approaches a constant. The solid red line shows the position of the front~$y$ as a function of time~$t$ obtained from the numerical solutions of Eq.~(4) in the main text. The dashed black line is the best linear fit to the data for~$t>5$. The position of the front is defined as~$y=\int_{0}^{\infty} c_{1}(x',y')dy'$. Here, the parameters are as in figure~5 of the main text, but only the first strain is present, and the habitat size is~$1\times10$ with the discretization step of~$2^{-7}$.}
\label{fig:velocity}
\end{figure}

\begin{figure}
\includegraphics[width=6cm]{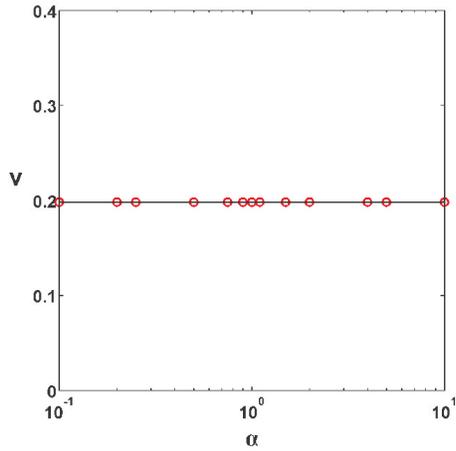} 
\caption{The velocity of colony expansion~$v$ does not depend on how fast the diffusion constant decreases with cell concentration. The red circles show the expansion velocities obtained from the numerical solutions of Eq.~(4) in the main text. The solid black line is a horizontal line indicating the absence of dependence of~$v$ on~$\alpha$; because only one strain is present we omit the indexes of~$v$ and~$\alpha$. Here, the parameters are as in figure~5 of the main text, but only the first strain is present, and the habitat size is~$1\times10$ with the discretization step of~$2^{-7}$.}
\label{fig:alpha}
\end{figure}

\begin{figure}
\includegraphics[width=6cm]{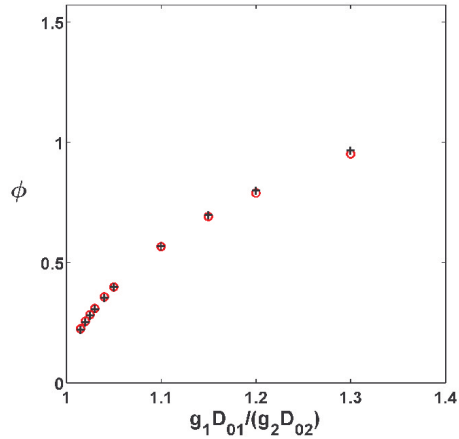} 
\caption{The sector angle~$\phi$ depends only on~$g_{1}D_{01}/(g_{2}D_{02})$. Equation~(4) in the main text is solved numerically for linear expansions. The resulting sector angles are plotted for varying~$g_{1}$~(shown with black plus signs) and varying~$D_{01}$~(shown with red circles). Here, the non-varying parameters are as in figure~5 of the main text, but the habitat size is~$1\times8$ with the discretization step of~$2^{-7}$.}
\label{fig:growth_diffusion}
\end{figure}

\begin{figure}
\includegraphics[width=6cm]{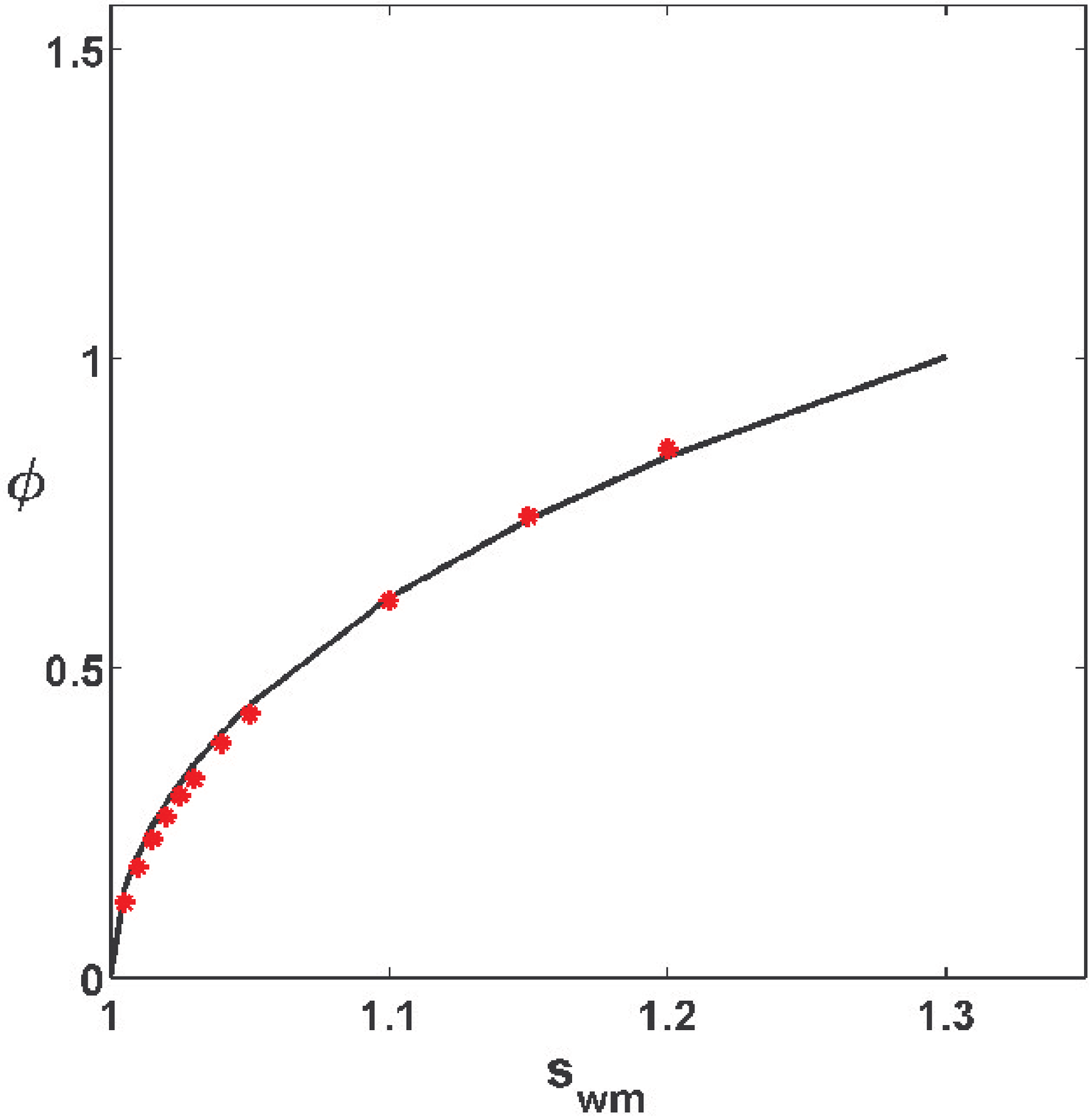} 
\caption{Non-monotonic dependence of diffusion constant on the cell concentration does not affect our results. Equation~(4) in the main text was solved numerically for linear expansions with~$D_{1}(c_{1},c_{2})=D_{2}(c_{1},c_{2})=D_{0}(c_{1}+c_{2})(1-c_{1}-c_{2})\theta(1-c_{1}-c_{2})$. The resulting sector angles~$\phi$ are shown with red stars and the theoretical expectation~$\cos(\phi/2)=(1+s_{\rm{wm}})^{-1/2}$ from equation (8) in the main text is shown with a solid black line. Note that, although equation (7) of the main text no longer holds,~$v^{2}_{1}/v^{2}_{2}=g_{1}D_{01}/(g_{2}D_{02})$ is still valid from dimensional analysis (see also Ref.~\protect{\cite{Murray:MathematicalBiology}} for a discussion of wave speeds when~$D\sim c$). Here, the non-varying parameters are as in figure~5 of the main text, but the habitat size is~$1\times8$ with the discretization step of~$2^{-7}$.}
\label{fig:diffusion_c}
\end{figure}

We first show that colony expansion can indeed be described by a constant velocity, see figure~\ref{fig:velocity}. The initial transient is due to the front relaxation from the sharp, step-like initial condition. We then demonstrate in figure~\ref{fig:alpha} that this velocity is insensitive to the exponents $\alpha_{1}$ and~$\alpha_{2}$, which control how fast migration stops behind the front. In figure~\ref{fig:growth_diffusion}, we show that the angle~$\phi$ describing sectors in the linear geometries does not depend on whether the beneficial mutation is caused by an increase in the growth rate or in the diffusion constant. Indeed, in agreement with equations~(7) and~(8) of the main text we find that~$\phi$ depends only on~$v_{1}/v_{2}=\sqrt{g_{1}D_{01}/(g_{2}D_{02})}$. Finally, we show in figure~\ref{fig:diffusion_c} that our results still hold even when the diffusion constant changes with cell concentration non-monotonically. In particular, the sector angle is still given by equation~(8) in the main text.

The results of our numerical exploration suggest that the macroscopic spatial patterns are determined only by~$v_{1}/v_{2}$. Therefore, there is some freedom in the choice of microscopic parameters, if one would like to use numerical solutions of equation~(4) in the main text to compute spatial patterns. In particular, we can set~$\alpha_{1}=\alpha_{2}=1$ and~$\epsilon_{1}=\epsilon_{2}=0$ for simplicity. The remaining four parameters~$g_{1}$,~$g_{2}$,~$D_{01}$, and~$D_{02}$ determine three observables: the time scale~$g_{1}^{-1}$, length scale~$\sqrt{D_{01}/g_{1}}$, and selective advantage~$(1+s)^{2}=g_{1}D_{01}/(g_{2}D_{01})$. Thus, there is one degree of freedom remaining, which we can fix if the relationship between selective advantage in the Petri dish~$s=1-v_{1}/v_{2}$ and in liquid culture~$s_{\rm{wm}}=g_{1}/g_{2}-1$ is known. In general, this relationship can be arbitrary because mutations can affect migration rate represented by~$D_{01}$ independently from the growth rate. However, two special cases deserve special attention.


The first special case assumes that~$D_{01}/D_{02}=g_{1}/g_{2}=1+s_{\rm{wm}}$, as is appropriate for compact colonies where migration is driven by cell growth and division. Equation~(4) in the main text then takes the following form

\begin{equation}
\left\{
\eqalign{
\frac{\partial c_{1}(t,\bm{x})}{\partial t}&=(1+s_{\rm{wm}})\bm{\nabla}\cdot\left[D(c_{1},c_{2})\bm{\nabla}c_{1}(t,\bm{x})\right]+ \\ & (1+s_{\rm{wm}})gc_{1}(1-c_{1}-c_{2}),\\
\frac{\partial c_{2}(t,\bm{x})}{\partial t}&=\bm{\nabla}\cdot\left[D(c_{1},c_{2})\bm{\nabla}c_{2}(t,\bm{x})\right]+\\ & gc_{2}(1-c_{1}-c_{2}),\\}
\right.
\label{ECompetitionSpatialSimplified}
\end{equation}

\noindent where~$g$ is the growth rate of strain~$2$, and~$D$ is its diffusion constant with the following dependence on the concentrations of cells

\begin{equation}
\label{SCDiffusionConstantSimplified}
D(c_{1},c_{2})=D_{0}(1-c_{1}-c_{2})\theta(1-c_{1}-c_{2}). 
\end{equation}

\noindent Under these assumptions, the selective advantage in liquid an in the Petri dish are the same~$s=s_{\rm{wm}}$.

The second special case is appropriate for growth-independent migration, e.g. swimming or swarming. Here~$D_{01}=D_{02}$, and equation~(4) in the main text takes the following form.

\begin{equation}
\left\{
\eqalign{
\frac{\partial c_{1}(t,\bm{x})}{\partial t}&=\bm{\nabla}\cdot\left[D(c_{1},c_{2})\bm{\nabla}c_{1}(t,\bm{x})\right]+ \\ & (1+s_{\rm{wm}})gc_{1}(1-c_{1}-c_{2}),\\
\frac{\partial c_{2}(t,\bm{x})}{\partial t}&=\bm{\nabla}\cdot\left[D(c_{1},c_{2})\bm{\nabla}c_{2}(t,\bm{x})\right]+\\ & gc_{2}(1-c_{1}-c_{2}),\\}
\right.
\label{ECompetitionSpatialSimplified2}
\end{equation}

\noindent Under these assumptions,~$1+s=\sqrt{1+s_{wm}}$. 

As we show in the main text, the first special case is more appropriate for yeast colonies. Note, however, that equation~(\ref{ECompetitionSpatialSimplified}) yields the same spatial patterns as equation~(\ref{ECompetitionSpatialSimplified2}) provided provided~$(1+s_{\rm{wm}})\to(1+s_{\rm{wm}})^{2}$.

\section{Duration of the initial stage of sector formation}\label{SSectorEstablishment}

In section 3 of the main text we discuss that sector formation proceeds in two stages: During the late stage, when the sector is large, its interior is occupied by the fitter strain only. The sector boundaries are far apart and are well described by the equal-time argument of section 4. During the early stage, the sector size is comparable to the width of the sector boundaries, and the two boundaries interact. In this regime, the sector interior is occupied by a mix of the two strain; the fitter strain has not yet completed the selective sweep. This effect is shown in simulations in figure 5 of the main text, and is also clearly visible in the experimental sectors shown in figure 1, 15 and 16 of the main text, as well as in figure \ref{fig:budding_mutants}(b).

In this early regime of sector establishment, the equal-time argument does not apply because the assumption that the domain of one strain is impenetrable to the other strain is violated. Our reaction-diffusion model predicts that the characteristic time duration of the early regime of sector formation scales as the inverse difference in growth rates of the two strains, $(g_1-g_2)^{-1}$, see equation (6) of the main text. This measn that the sector establishment time should decrease with the fitness difference of the two strains.

\begin{figure}
\includegraphics[width=10cm]{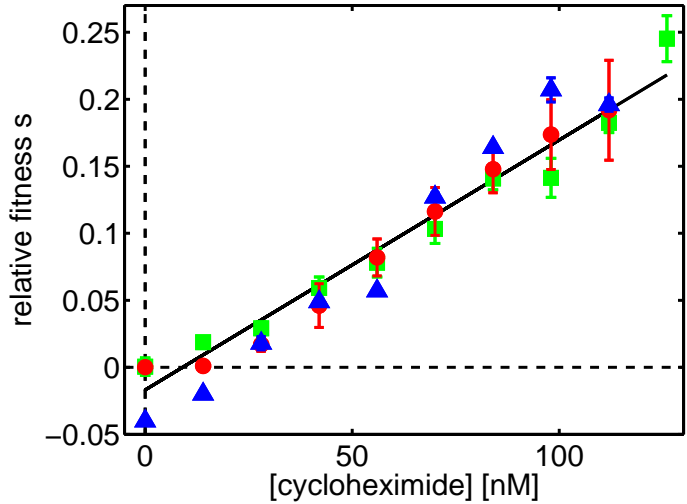} 
\caption{Dependence of the relative fitness $s$ on the drug cycloheximide, as measured by a liquid fitness assay (blue triangles), by radial expansion sectors (red circles), and by colony collisions (green squares). The solid black line is a fit to all data points.}
\label{fig:Fitness_CHX}
\end{figure}

\begin{figure}
\includegraphics[width=7.5cm]{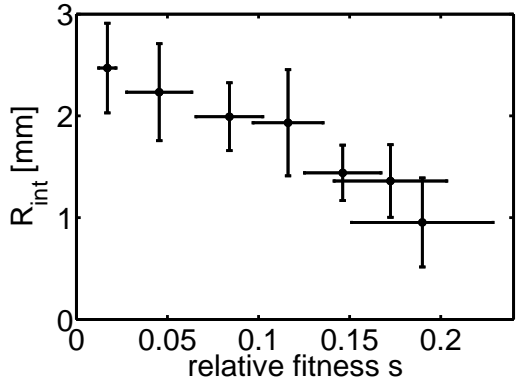} 
\caption{Characterization of the sector establishment: The distance $R_{\rm int}$ between the homeland and the intersection of the fits to the sector arms, determined from the same experiments as in figure~\ref{fig:Fitness_CHX}, decreases with the relative fitness $s$.}
\label{fig:Sector_establishment}
\end{figure}

In order to test this experimentally, we varied relative fitness using the drug cycloheximide for the competition of the cycloheximide-sensitive wild-type strain yJHK112 and the cycloheximide-resistant strain yMM8. We then measured the relative fitness by our three most accurate direct competition methods: the liquid fitness assay, the radial expansion sector assay, and the colony collision assay. Figure~\ref{fig:Fitness_CHX} shows that the results from the three methods agree reasonably well with each other, and that the relative fitness increases linearly with the cycloheximide concentration. 

Experimentally, the sector establishment time can be characterized by the distance $R_{\rm int}$  between the homeland and the point of intersection of the of the two tangents fitted to the sector in the late, fully established, stage, see figures 15(b) and 16(b) of the main text. We determined $R_{\rm int}$ for the radial expansion sectors for competitions at different cycloheximide concentrations and thus for different fitnesses $s$, see figure~\ref{fig:Sector_establishment}. 
For fitnesses larger than zero\footnote{We omitted sectors with fitness values indistinguishable from zero, since we could not fit these sectors with our usual stringent criterion, see section~\ref{SExperiments}. In addition, our theoretical prediction in equation (6) of the main text breaks down for zero fitness differences.}, the distance $R_{\rm int}$ in decreases with relative fitness $s$. This means that indeed the sector establishment time decreases for increasing fitness.

\section{Statistical analysis of differences between fitness assays}\label{SStatisticalAnalysis}

In this paper, we measure the relative fitness of two yeast strains using different spatial assays on agar plates as well as the standard fitness assays in liquid culture. Here, we compare the results of different fitness assays, and investigate whether they are significantly different. 
To this purpose, we apply statistical hypothesis testing with a significance level of $0.05$. When performing multiple comparisons in parallel, we use a Bonferroni correction: For $n$ comparisons, each test is required to have a $p$-value smaller than $0.05/n$ to give an overall significance level of $\sim 0.05$ \cite{Shaffer95}.

Table 1 in the main text summarizes the results of all different fitness assays for the competition of the advantageous sterile mutant and the yellow reference strain. Inspection of this table immediately shows that some assays have a very high standard deviation, namely the two linear expansion assays and the radial expansion velocity assay. Presumably due to this high variability, the results from these assays cannot be significantly distinguished from each other, nor from the other assays shown in the table: $p$-values from Welch's t-test, which tests whether two data sets have equal means (but allowing for unequal variances \cite{Weiss07}), are all larger than $0.25$. Thus we cannot reject the hypothesis that these assays would give the same result, given sufficient amount of data.

We now focus on the more accurate fitness assays: the radial expansion sectors and colony collisions, as well as the liquid culture competition assay (the liquid culture growth rate assay behaved similar to the liquid competition assay). We first used a One-way ANOVA F-test to test whether these three assays had the same means \cite{Weiss07}. This was not the case ($p=0.008$). Pairwise comparison of the three assays with Welch's t-test showed that only the radial sector assay was significantly different ($p<0.05/3$, using a Bonferroni correction for 3 comparisons) from the liquid culture competition, as well as from the colony collision assay.  

We next applied the same statistical testing procedure to the competitions of the budding mutant and the advantageous sterile mutant whose results are shown in table~\ref{table:budding_mutants}. Again the three assays do not have the same means ($p=0.008$ from an ANOVA F-test), but for this competition the liquid competition result was different from both the radial sector assay and the colony collision assay ($p<0.05/3$), while the two plate assays were not significantly different.

We also applied our testing procedure to the cycloheximide experiment described in section~\ref{SSectorEstablishment}, in which the liquid culture competition assay, the radial sector assay, and the collision assay are employed over a range of fitnesses, see figure~\ref{fig:Fitness_CHX}. One-way ANOVA indicated significant differences ($p<0.05$) for the three lowest cycloheximide concentrations, as well as for 98nM. Pairwise comparisons with Welch's t-test showed that the liquid culture competition was significantly different from the two spatial assays for 0, 14 and 98 nM, while the two spatial assays were different for 14 and 28nM. 

\begin{table}
\begin{tabular}{|l@{\;\;}|l@{\;\;}|l@{\;\;}|}\hline
Assays 									& mean fitness difference\\ \hline
Radial sectors and liquid competition	& \phantom{-}0.002 $\pm$ 0.023\\
Collision and liquid competition		& \phantom{-}0.000 $\pm$ 0.034\\
Radial sectors and collisions			& \phantom{-}0.002 $\pm$ 0.015\\\hline 
\end{tabular}
\caption{Average differences of relative fitnesses determined by different assays for the cycloheximide data shown in figure~\ref{fig:Fitness_CHX}. Errors are standard deviations.}
\label{tab:fitness_differences}
\end{table}

The differences between the different fitness assays could be caused by a systematic error in sector and collision analysis, or by the underestimation of measurement variance due to possible data clustering by the day of the experiment. In addition, the three-dimensional structure of yeast colonies (yeast colonies grow to a height of about 1mm), which is neglected in our two-dimensional theory, might cause some of the differences between the spatial and liquid assays.

To get an overall impression of the assay performance, we calculated the mean differences of relative fitnesses determined by different assays for the cycloheximide data, see table~\ref{tab:fitness_differences}. The average differences are very close to zero, indicating that the assays give similar results without a directional bias. The standard deviations are of the order of $0.02$, suggesting that the assays might give different results for fitness differences of this order of magnitude. Indeed, in the cycloheximide experiment, the assays are significantly different for fitness values $s<0.02$, as described above, but not for larger fitnesses except one. 

In summary, we conclude that the three relatively accurate fitness assays, namely the liquid competition assay and the spatial radial sector and colony collision assays give overall similar results, especially for large fitness differences.


\section*{References}

\bibliography{SCBibs}
\bibliographystyle{unsrt}

\end{document}